\documentclass[manuscript]{aastex}

\slugcomment{to be submitted for the regular issue of the Astrophysical Journal}

\shorttitle{Precursors of the Forbush Decrease }
\shortauthors{Fushishita et al.}

\begin{document}


\title{Precursors of the Forbush Decrease on December 14, 2006 observed with the Global Muon Detector Network (GMDN)}


\author{
 A. Fushishita\altaffilmark{1},
 T. Kuwabara\altaffilmark{2},
 C. Kato\altaffilmark{1},
 S. Yasue\altaffilmark{1},
 J. W. Bieber\altaffilmark{2},
 P. Evenson\altaffilmark{2},
 M. R. Da Silva\altaffilmark{3},
 A. Dal Lago\altaffilmark{3},
 N. J. Schuch\altaffilmark{4},
 M. Tokumaru\altaffilmark{5},
 M. L. Duldig\altaffilmark{6},
 J. E. Humble\altaffilmark{7},
 I. Sabbah\altaffilmark{8,9},
 H. K. Al Jassar\altaffilmark{10},
 M. M. Sharma\altaffilmark{10}, and
 K. Munakata\altaffilmark{1} 
}


\altaffiltext{1}{Physics Department, Shinshu University, Matsumoto, Nagano 390-8621, Japan}
\altaffiltext{2}{Bartol Research Institute and Department of Physics and Astronomy, University of Delaware, Newark, DE 19716, USA}
\altaffiltext{3}{National Institute for Space Research (INPE), 12227-010 Sao Jose dos Campos, SP, Brazil}
\altaffiltext{4}{Southern Regional Space Research Center (CRS/INPE), P.O. Box 5021, 97110-970, Santa Maria, RS, Brazil}
\altaffiltext{5}{Solar Terrestrial Environment Laboratory, Nagoya University, Nagoya, Aichi 464-8601, Japan}
\altaffiltext{6}{Australian Antarctic Division, Kingston, Tasmania 7050, Australia}
\altaffiltext{7}{School of Mathematics and Physics, University of Tasmania, Hobart, Tasmania 7001, Australia}
\altaffiltext{8}{Astronomy Department, Faculty of Science, King Abdulaziz University, Jeddah  Saudi Arabia}
\altaffiltext{9}{On leave from Department of Physics, Faculty of Science, Alexandria University, Alexandria, Egypt}
\altaffiltext{10}{Physics Department, Faculty of Science, Kuwait University, Kuwait City, Kuwait}


\begin{abstract}
We analyze the precursor of a Forbush Decrease (FD) observed with the Global Muon Detector Network on December 14, 2006. An intense geomagnetic storm is also recorded during this FD with the peak Kp index of 8+. By using the ``two-dimensional map'' of the cosmic ray intensity produced after removing the contribution from the diurnal anisotropy, we succeed in extracting clear signatures of the precursor. A striking feature of this event is that a weak loss-cone signature is first recorded more than a day prior to the Storm Sudden Commencement (SSC) onset. This suggests that the loss-cone precursor appeared only 7 hours after the Coronal Mass Ejection (CME) eruption from the Sun, when the Interplanetary (IP) shock driven by the Interplanetary Coronal Mass Ejection (ICME) located at 0.4 AU from the Sun. We find the precursor being successively observed with multiple detectors in the network according to the Earth's spin and confirmed that the precursor continuously exists in space. The long lead time (15.6 hours) of this precursor which is almost twice the typical value indicates that the IMF was more quiet in this event than a typical power spectrum assumed for the IMF turbulence. The amplitude (-6.45 \%) of the loss-cone anisotropy at the SSC onset is more than twice the FD size, indicating that the maximum intensity depression behind the IP shock is much larger than the FD size recorded at the Earth in this event. We also find the excess intensity from the sunward IMF direction clearly observed during $\sim$10 hours preceding the SSC onset. It is shown that this excess intensity is consistent with the measurement of the particles accelerated by the head-on collisions with the approaching shock. This is the first detailed observation of the precursor due to the shock reflected particles with muon detectors. 
\end{abstract}





\section{Introduction}

A solar disturbance propagating away from the Sun affects the pre-existing population of galactic cosmic rays (GCRs) in a number of ways. Most well-known is the ``Forbush Decrease'' (FD), a region of suppressed cosmic-ray density located downstream of an Interplanetary Coronal Mass Ejection (ICME) shock. Some particles from this region of suppressed density leak into the upstream region and, traveling nearly at the speed of light, they race ahead of the approaching shock and are observable as a precursory loss cone (LC) anisotropy far into the upstream region \citep{Barnden71} \citep{Nagashima92}. LCs are characterized by intensity deficits confined to a narrow pitch-angle region around the sunward direction along the Interplanetary Magnetic Field (IMF) and are typically visible 4-8 hours ahead of shock arrival associated with major geomagnetic storms \citep{kmuna00}.

On the basis of numerical simulations of the high-energy particle transport across the shock, \citet{Lee03}[hereafter referred to as paper 1] derived the theoretical constraints for the LC anisotropy. In particular, they presented the quantitative relationship between the angular width of the pitch-angle distribution and the interplanetary parameters which include the angle between the shock normal and the upstream IMF and the mean free path of the pitch-angle scattering of GCRs in the turbulent magnetic field. By analyzing the ``two-dimensional map'' of the cosmic ray intensity observed with a large single muon hodoscope during a LC precursor period in October, 2003, \citet{kmuna05} reported a lead time of 4.9 hours for the LC precursor indicating a mean free path of the pitch-angle scattering shorter than the theoretical expectation based on the numerical simulation. This implies that the IMF was more turbulent in the event analyzed by them, than a typical power spectrum of the IMF turbulence assumed in paper 1. They also reported, on the other hand, a rather broad pitch-angle distribution for the event, which implies a ``quasi-parallel'' shock according to the numerical simulation. Although this seemed to be conflicting with in-situ IMF and plasma data suggesting a ``quasi-perpendicular'' shock, it was discussed that this apparent conflict can be resolved by taking account of a pair of shocks formed by ICMEs successively ejected from the Sun. The ``Alaska model'' simulation of the event indicated that a large shock responsible for the SSC at the Earth was overtaking another smaller shock westward of the Earth. During the loss cone period, therefore, it was possible that the Earth may have been connected with the westward shock which was more consistent with a quasi-parallel geometry. Thus, the situation of this event was rather complicated introducing an additional ambiguity in the comparison of the observation with theoretical predictions. In this paper, we analyze another simpler LC event associated with a single shock.

Cosmic ray precursors observed with a long lead time is of particular importance for the possible space weather forecast using cosmic ray measurements. An accurate measurement of such a precursor, however, requires detectors in a global network for continuous monitoring over a long time period more than a day prior to the SSC onset \citep{kmuna00,Belov01}. Such a network has been realized only recently by the Global Muon Detector Network (GMDN) \citep{Okazaki08}. For accurate analyses of LC events, it is also necessary to properly remove the contribution from the diurnal anisotropy, which always exists in space with an amplitude comparable to the intensity deficit due to the LC anisotropy. The GMDN can also measure precisely the diurnal anisotropy utilizing the global sky coverage of the network \citep{Okazaki08}. In this paper, we develop new analysis methods for removing the influence of the diurnal anisotropy and apply the ``two-dimensional map'' analysis by \citet{kmuna05} to the cosmic-ray precursors observed with the GMDN in December 2006 for better visualization of the signatures of cosmic ray precursors.

The outline of this paper is as follows. In \S 2, we briefly summarize the features of the event which we analyze in this paper. In \S 3, we apply a new analysis method to the data observed by the GMDN and present the best-fit analyses of a model based on the theoretical prediction by paper 1. Our conclusions and discussions based on the comparison with the theoretical prediction are given in \S 4.

\section{Event overview}

An X3.4 flare was first observed at 02:14 UT on December 13, 2006 and was followed by a halo Coronal Mass Ejection (CME) eruption shortly after at 02:54 UT. This CME was accompanied by a strong interplanetary (IP) shock, which traveled interplanetary space with an average velocity of 1160 km/s and arrived at the Earth causing the Storm Sudden Commencement (SSC) onset at 14:14 UT on December 14 \citep{Liu08}. The shock speed at 1 AU was 1030 km/s as calculated from the conservation of the plasma mass across the shock. It was also derived from the Rankine-Hugoniot relations \citep{Vinas86} that the shock normal makes an angle of about 56$\degr$ with the upstream magnetic field indicating the quasi-perpendicular shock. An intense geomagnetic storm followed the SSC with peak Kp index of 8+. There was no other SSC and interplanetary disturbances recorded at the Earth between 02:14 UT on December 13 and 14:14 UT on December 14. From the data recorded by the GMDN, we routinely derive the anisotropy vector (or the first order anisotropy) by fitting the function $I_{i,j}(t)$ given by 
\begin{eqnarray}
I_{i,j}(t) = I^{0}_{i,j}(t)&+&\xi^{GEO}_x(t)(c^{1}_{1i,j} \cos \omega t_i - s^{1}_{1i,j}\sin \omega t_i)
\nonumber \\
  &+& \xi^{GEO}_y(t)(s^{1}_{1i,j}\cos \omega t_i + c^{1}_{1i,j}\sin \omega t_i) \nonumber \\
  &+& \xi^{GEO}_z(t)c^{0}_{1i,j} \label{eq01}
\end{eqnarray}
to the pressure-corrected hourly count rate $I^{obs}_{i,j}(t)$ at universal time $t$ in the $j$-th directional channel of the $i$-th muon detector in the GMDN. In this equation, $I^{0}_{i,j}(t)$ denotes the contribution from the cosmic ray density $I_{0}(t)$ and $\xi^{GEO}_{x}(t)$, $\xi^{GEO}_{y}(t)$, $\xi^{GEO}_{z}(t)$ are the best-fit parameters denoting three components of the anisotropy vector in the local geographical coordinate system (GEO), $c^{1}_{1i,j}$, $s^{1}_{1i,j}$ and $c^{0}_{1i,j}$ are the coupling coefficients calculated by assuming a rigidity independent anisotropy, $t_i$ is the local time at the location of the $i$-th detector and $\omega = \pi /12$ \citep[for more detail of our analyses, see][]{Okazaki08}. Figs.1a-1b display the hourly data of the solar wind velocity and the IMF magnitude measured by the {\it ACE} satellite over three day period between December 13 and 15, each as a function of time in the day of year (DOY) on the horizontal axis. The left and right vertical solid lines indicate the flare onset and the SSC onset, respectively. A clear signature of the IP shock is seen in abrupt increases of the solar wind velocity and the IMF magnitude following the SSC onset. The IMF data observed during the period between the flare and SSC onsets show no signature of the IMF sector boundary and indicate that the Earth was in the ``toward'' sector throughout the period. Figure 1c shows hourly values of the derived cosmic ray density ($I_{0}(t)$), whereas Figs.1d-1f display three anisotropy components ($\xi^{GSE}_{x}(t)$, $\xi^{GSE}_{y}(t)$, $\xi^{GSE}_{z}(t)$) transformed to the geocentric solar ecliptic (GSE) coordinate system. The anisotropy components shown in Figs.1d-1f are corrected for both the solar wind convection anisotropy and the Compton-Getting anisotropy arising from the Earth's 30 km/s orbital motion around the Sun. In Figure 1c, a FD is clearly observed with a $\sim$3 \% maximum depression of the cosmic ray density following the onset of the SSC as indicated by the vertical solid line. The average feature of the anisotropy components over the period preceding the SSC in Figs.1d-1f indicates $\xi^{GSE}_{x}(t)<0$, $\xi^{GSE}_{y}(t)>0$ and $\xi^{GSE}_{z}(t)>0$, which is in a qualitative agreement with the drift model prediction for the anisotropy in the ``toward'' IMF sector during the $A<0$ epoch of the solar polar magnetic field \citep{Okazaki08, Fushishita10}.

Also plotted by black points in Figure 1g is the pressure-corrected count rate recorded in the vertical channel of the S\~{a}o Martinho da Serra detector in Brazil (hereafter S\~{a}o Martinho). It is evident in this panel that S\~{a}o Martinho observed a sharp intensity decrease around DOY 348.4 a few hours prior to the SSC onset indicating the LC precursor. We note in this figure that $I_{0}(t)$, $\xi^{GSE}_{x}(t)$, $\xi^{GSE}_{y}(t)$ and $\xi^{GSE}_{z}(t)$ are all changing in response to the LC precursor recorded by the S\~{a}o Martinho detector. As shown later, the residual of the best-fitting is larger in two periods of 347.563- 347.771 and DOYs 348.271-348.563, with the second period containing the LC signature in S\~{a}o Martinho data in Figure 1g. This shows that the simple model in eq. (\ref{eq01}) cannot properly describe the LC anisotropy. In the next section, we improve this model by taking account of the contribution from the LC anisotropy.

By using the {\it ACE} IMF data, we first examine the asymptotic viewing direction along the sunward IMF in 24 hours prior to the SSC onset in December 14, together with viewing directions (after correction for geomagnetic bending) available in the GMDN. Although there is a large fluctuation in the IMF orientation observed by {\it ACE}, we find that the majority of the IMF orientation is in the southern hemisphere, as expected for the sunward direction along the nominal Parker field in the ecliptic plane viewed from the Earth in December. The LC precursor, therefore, is expected to be observed in the directional channels viewing around the equator or a little south of the equator. This is consistent with the LC signature observed by S\~{a}o Martinho in the bottom panel of Figure 1. Our detailed analysis in the next section will show the LC precursor also observed by a small detector at Hobart viewing the mid-latitude in the southern hemisphere. In the following analysis, we use for calculating the pitch-angle of each viewing direction in the GMDN the nominal Parker field calculated from the solar wind velocity by {\it ACE} \citep{Parker58}, instead of using the {\it ACE} IMF, to avoid effects of the large fluctuation in the {\it ACE} IMF orientation. The anisotropy of $\sim$30 GeV GCR intensity observed with muon detectors is rather stable, changing only gradually even when the {\it ACE} IMF shows a large fluctuation. 

\section{Data analysis and results}

Muon detectors measure high-energy GCRs by detecting secondary muons produced from the hadronic interactions of primary GCRs (mostly protons) with atmospheric nuclei. Since relativistic muons have relatively long lifetimes (the proper half-life being 2.2 $\mu$s) and can reach the ground preserving the incident direction of the initiating primary particles, we can measure the GCR intensity in various directions with a multi-directional detector at a single location. The typical energy of primary GCRs modulated by FDs is $\sim$30 GeV for muon detectors. The Global Muon Detector Network (GMDN) originally started with three detectors at Nagoya (Japan), Hobart (Australia) and S\~{a}o Martinho da Serra (Brazil), each of which is multi-directional, thus allowing us to simultaneously record the muon intensities in various directions of viewing. These detectors have identical design, except for their detection area which is 36 m$^2$ for Nagoya, 9 m$^2$ for Hobart and 28 m$^2$ for S\~{a}o Martinho, consisting of two horizontal layers of plastic scintillators, vertically separated by 1.73 m, with an intermediate 5 cm layer of lead to absorb the soft component radiation in the air. Each layer consists of an array of 1 m$^2$ unit detectors, each with a 1 m$\times$1 m plastic scintillator viewed by a photomultiplier tube of 12.7 cm diameter. By counting pulses of the 2-fold coincidences between a pair of detectors on the upper and lower layers, we can record the rate of muons from the corresponding incident direction. The multi-directional muon telescope comprises various combinations between the upper and lower detectors. The Field of View (FOV) of the vertical channel has a broad geometrical aperture of $\sim\pm30^\circ$ ($\pm \tan^{-1}(1.0/1.73)$). A precise measurement of LCs using muon detectors became possible only recently, when the GMDN capable of continuously monitoring the sunward IMF direction was completed in March 2006 by adding another 9 m$^2$ detector at Kuwait University. Unlike the other three detectors, the Kuwait Muon Telescope (hereafter Kuwait) consists of four horizontal layers of 30 proportional counter tubes (PCTs). Each PCT is a 5 m long cylinder with a 10 cm diameter having a 50-micron thick tungsten anode along the cylinder axis. A 5 cm layer of lead is installed above the detector to absorb the soft component radiation. The PCT axes are aligned geographic east-west (X) in the top and third layers and north-south (Y) in the second and bottom layers. The top and second layers form an upper pair, while the third and bottom layers form a lower pair. The two pairs are separated vertically by 80 cm. Muon recording is triggered by the fourfold coincidence of pulses from all layers and the incident direction is identified from X-Y locations of the upper and lower PCT pairs. This is approximately equivalent to recording muons with two 30$\times$30 square arrays of 10 cm$\times$10 cm detectors separated vertically by 80 cm. The geometrical aperture of the FOV in this detector is $\sim\pm7^\circ$ ($\pm\tan^{-1}(0.1/0.8)$) for the vertical channel. The muon count is recorded in each of 23$\times$23=529 directional channels which cover 360$\degr$ of azimuth angle and 0$\degr$ to 60$\degr$ zenith \citep[for more detail of GMDN, see][]{Okazaki08}.

The number of viewing directions available from the GMDN was drastically increased from the conventional one by installing a new recording system using the Field Programmable Gate Arrays (FPGAs), with which we can count muons for all possible coincidences between a pair of unit detectors on the upper and lower layers \citep{Yasue03}. If we have a square $n \times n$ array of unit detectors aligned to the north-south (or east-west) direction in the $i$-th muon detector, it is possible to analyze the pressure corrected muon rates $I^{obs}_{i(k,l)}(t) (k, l=-n+1,..., 0,...,n-1)$ in total $(2n-1)\times(2n-1)$ directional channels at the time $t$, where positive (negative) $k$ and $l$ represent eastern (western) and northern (southern) incidence, respectively, with $k=l=0$ corresponding to the vertical incident. In order to visualize the directional distribution of GCR intensity, we plot $I^{obs}_{i(k,l)}(t)$ as a function of $k$ and $l$ in a color-coded format, which we call the ``2D map'' (two dimensional map of the cosmic ray intensity). The number of directional channels used in this paper is 25 (5$\times$5) from Hobart, 121 (11$\times$11) from Kuwait and 49 (7$\times$7) from S\~{a}o Martinho, respectively. We cannot apply this technique to Nagoya data in this paper, as the FPGA recording system was installed in the Nagoya detector only in May, 2007 and not available for the FD event in December 2006. In this paper, therefore, we analyze only the conventional 17 direction data from Nagoya \citep[for the conventional directional channels in the GMDN, see][]{Okazaki08}. Thus, the total number of the directional channels analyzed in this paper is 212 from four detector systems in the GMDN.

\subsection{Analysis method and the sky-maps of the observed anisotropy}

As stated in \S 1, we need to accurately remove from the data the contribution from the diurnal anisotropy (DA) for precise analysis of the LC precursor. This was not possible for us before completing the GMDN with which we can precisely measure the DA utilizing the global sky coverage of the network. By using the 24-hour trailing moving averages (TMAs) of $I_0 (t)$, $\xi_x^{GEO} (t)$, $\xi_y^{GEO} (t)$ and $\xi_z^{GEO} (t)$ in eq. (\ref{eq01}), we calculate the contribution from the DA to $I_{i(k,l)}(t)$ at a universal time {\it t} as
\setlength{\arraycolsep}{0.0em}
\begin{eqnarray}
  \bar{I}_{i(k,l)}(t) = \bar{I}^{0}_{i(k,l)}(t)&+& \bar{\xi}^{GEO}_x(t)(c^{1}_{1 i(k,l)} \cos \omega t_i - s^{1}_{1 i(k,l)}\sin \omega t_i) \nonumber \\
  &+& \bar{\xi}^{GEO}_y(t)(s^{1}_{1 i(k,l)}\cos \omega t_i + c^{1}_{1 i(k,l)}\sin \omega t_i) \nonumber \\
  &+& \bar{\xi}^{GEO}_z(t)c^{0}_{1 i(k,l)}, \label{eq02}
\end{eqnarray}
where $\bar{I}^{0}_{i(k,l)}(t)$ is the 24-hour TMA of the contribution from the cosmic ray density ($I_{0}(t)$) and $\bar{I}_{0}(t)$, $\bar{\xi}^{GEO}_x(t)$, $\bar{\xi}^{GEO}_y(t)$ and $\bar{\xi}^{GEO}_z(t)$ are the 24-hour TMAs of the best-fit parameters in eq. (\ref{eq01}) calculated as
\setlength{\arraycolsep}{0.0em}
\begin{eqnarray}
  \bar{I}_{0}(t)       &=& \Sigma^t_{t-23} I_{0}(t)/24,       \label{eq03} \\
  \bar{\xi}^{GEO}_x(t) &=& \Sigma^t_{t-23} \xi^{GEO}_x(t)/24, \label{eq04} \\
  \bar{\xi}^{GEO}_y(t) &=& \Sigma^t_{t-23} \xi^{GEO}_y(t)/24, \label{eq05} \\
  \bar{\xi}^{GEO}_z(t) &=& \Sigma^t_{t-23} \xi^{GEO}_z(t)/24. \label{eq06}
\end{eqnarray}
By subtracting $\bar{I}_{i(k,l)}(t)$ in eq. (\ref{eq02}) from the observed $I^{obs}_{i(k,l)}(t)$, we get the directional intensity distribution $\Delta I^{obs}_{i(k,l)}(t)$ free from the DA as
\begin{equation}
  \Delta I^{obs}_{i(k,l)}(t) = I^{obs}_{i(k,l)}(t)-\bar{I}_{i(k,l)}(t).\label{eq07}
\end{equation}
We cannot use $I_0 (t)$, $\xi_x^{GEO} (t)$, $\xi_y^{GEO} (t)$, $\xi_z^{GEO} (t)$ in eq. (\ref{eq02}) instead of $\bar{I}_{0}(t)$, $\bar{\xi}^{GEO}_x(t)$, $\bar{\xi}^{GEO}_y(t)$, $\bar{\xi}^{GEO}_z(t)$, respectively, because the LC signature recorded in a large detector like S\~{a}o Martinho makes the best-fit $I_{i(k,l)}(t)$ too close to the observed value and consequently leads to too small $\Delta I^{obs}_{i(k,l)}(t)$ in eq. (\ref{eq07}). In Figure 1g, $I^{obs}_{i(k,l)}(t)$, $\bar{I}_{i(k,l)}(t)$ and $\Delta I^{obs}_{i(k,l)}(t)$ for the vertical channel of S\~{a}o Martinho are displayed by solid circles, a gray curve and open circles, respectively. It is seen that the gradual variation of $I^{obs}_{i(k,l)}(t)$ due to the DA ($\bar{I}_{i(k,l)} (t)$) is successfully removed in $\Delta I^{obs}_{i(k,l)}(t)$. We have calculated using 8-hour and 12-hour TMAs for $\bar{I}_{i(k,l)}(t)$ instead of 24-hour TMA and have confirmed that the results remain essentially unchanged. Note that $\Delta I^{obs}_{i(k,l)}(t)$ in eq. (\ref{eq07}) is derived using the ``trailing'' average and is not affected by the variation occurring after $t$. This is an important issue for possible real time forecasting.

To visualize the LC signature more clearly by suppressing statistical fluctuation which is larger in the inclined channels, we also use, instead of $\Delta I^{obs}_{i(k,l)}(t)$, the ``significance'' $s I_{i(k,l)}(t)$ defined as
\begin{equation}
  s I^{obs}_{i(k,l)}(t) = \Delta I^{obs}_{i(k,l)}(t)/\sigma_{i(k,l)}, \label{eq08}
\end{equation}
where $\sigma_{i(k,l)}$ is the count rate error for the $(k,l)$ directional channel in the $i$-th detector. Figure 2a shows the 2D maps of $s I^{obs}_{i(k,l)}(t)$ in eq. (\ref{eq08}) for S\~{a}o Martinho on December 13 and 14 during 36 hours preceding the SSC onset at 14:14 UT on December 14 (DOY 348.593), whereas Figure 2c shows $s I^{obs}_{i(k,l)}(t)$ for Hobart in the same period. Each small square panel in these figures displays $s I^{obs}_{i(k,l)}(t)$ observed in one hour in a color-coded format as a function of $k$ and $l$ on the horizontal and vertical axes denoting the east-west and north-south inclinations of the viewing direction, respectively. Red color denotes the excess intensity in each ($k,l$) pixel relative to the ominidirectional intensity in an entire FOV, while blue color denotes the deficit intensity. In these figures, we set color scales ranging $\pm$ 5 for S\~{a}o Martinho and $\pm$ 3 for Hobart taking account of larger statistical error in Hobart due to its smaller detection area. Also shown by the white curve in each panel is the contour line of the pitch-angle measured from the sunward IMF direction and calculated for cosmic rays incident to each ($k,l$) pixel with the median primary energy appropriate to that pixel (we assume $1/P$ rigidity spectrum for the LC anisotropy throughout this paper). It is seen that the zero pitch-angle region is first captured in the FOV by S\~{a}o Martinho in DOYs 347.354-347.563 (08:00-13:00 UT, December 13), and then by Hobart in DOYs 347.771-347.979 (18:00-23:00 UT, December 13) and by S\~{a}o Martinho again in DOYs 348.354-348.563 (08:00-13:00 UT, December 14) according to the Earth's spin. Hereafter, these three 6-hour periods are referred to as the LC periods. In Figure 2a for S\~{a}o Martinho, there is a clear LC precursor seen as a deficit intensity at around $0^{\circ}$ pitch-angle with the minimum intensity about -0.8 \% in DOYs 348.396$\sim$348.521 (09:00$\sim$12:00 UT on December 14) during one of the LC periods. A striking feature of this event is that a weaker LC signature is also seen one day earlier in DOYs 347.437$\sim$347.521 (10:00$\sim$12:00 UT on December 13) during another LC period. This suggests that the LC precursor already existed only 7 hours after the CME eruption at 02:54 UT on December 13, when the IP shock driven by an ICME located at 0.4 AU from the Sun \citep{Liu08}. If this is the case, the signature also should have been observed during a period in between these two LC periods by other detectors viewing the eastern sky neighboring the FOV of S\~{a}o Martinho. This is actually seen in $s I^{obs}_{i(k,l)}(t)$ for Hobart in Figure 2c in DOYs 347.771$\sim$348.021 (18:00, December 13$\sim$00:00 UT, December 14). In DOYs 348.188$\sim$348.271 (04:00$\sim$07:00 UT on December 14), a weak LC signature is also seen in Figure 2e for Kuwait viewing the sky in between the FOVs of S\~{a}o Martinho and Hobart. It is rather surprising that the LC signature passing along the southern edge of the FOV of Kuwait can be seen even with such a small detector (thanks to a better directional resolution of the muon hodoscope). We add to note that a large muon hodoscope GRAPES-3 in operation at Ooty in southern India also recorded a clear LC precursor on $\sim$09:00 LT (03:30 UT) in December 14, just before the signature recorded in Kuwait in Figure 2e \citep{Nonaka06}[Dr. H. Kojima, private communication]. These all give observational supports for the picture that the LC signature was continuously existing and was successively observed with the multiple detectors in the GMDN according to the Earth's spin.

There is also a clear intensity excess (EX) seen with the maximum intensity about +0.5 \% at $30^{\circ}$-$90^{\circ}$ pitch-angle particularly in DOYs 348.188-348.563 (04:00-13:00 UT on December 14) in Figs. 2a and 2c, partly overlapping the LC signature in Figure 2a for S\~{a}o Martinho. Such an intensity excess is expected from the ground based measurement of GCRs reflected by the IP shock approaching the Earth \citep{Dorman95, Belov95}. The numerical model for GCR transport across the shock also predicts such an EX anisotropy superposed on the LC distribution (paper 1). The fractional energy gain for a GCR traveling along the IMF with an energy $E$ after the reflection by the IP shock approaching with a velocity $V_{S}$ is calculated as
\begin{equation}
  \Delta E/E \approx 2V_{S} \cos \theta_{Bn}/c, \label{eq09}
\end{equation}
where $\theta_{Bn}$ is the angle between the shock normal and the upstream IMF and $c$ is the speed of light. The intensity excess expected from this energy gain is estimated as
\begin{equation}
  \Delta I/I = \gamma \Delta E/E \approx 2 \gamma V_{S} \cos \theta_{Bn}/c, \label{eq10}
\end{equation}
where $\gamma$ is the power-law index of the GCR energy spectrum, which we set to 2.7. By using the IP shock velocity (1030 km/s) at 1 AU \citep{Liu08} for $V_{S}$ and by tentatively assuming $\theta_{Bn}=60^{\circ} (30^{\circ})$, we get $\sim$+0.9 \%(+1.6 \%) for $\Delta I/I$.


We confirm that there is no notable excess or deficit intensity in $s I^{obs}_{i(k,l)}(t)$ except that in the periods mentioned above. In the following subsection, we will present detailed comparisons between the observation and the theoretical model for the LC and EX anisotropies.

\subsection{Model anisotropy and best-fitting to the observation}

Figure 3 shows the pitch-angle distribution of the GCR intensity. Open circles in this figure display a sample numerical pitch-angle distribution near an oblique interplanetary shock taken from Fig.5 of paper 1. In this figure, the GCR intensity is plotted as a function of the pitch-angle cosine ($\mu$). A sharp LC around $\mu=1$ (zero pitch-angle) and a broad excess intensity around $\mu=0.5$ ($60^{\circ}$ pitch-angle) are both evident in this figure. We model this distribution by a superposition of the LC and EX anisotropies, shown by plus signs and crosses in Figure 3, respectively, as given by
\begin{equation}
\hspace{0.5cm}
f(\theta,P,t)=C_{LC}(t)f_{LC}(\theta,P)+C_{EX}(t)(1+\cos \theta ), \label{eq11}
\end{equation}
where
\begin{equation}
f_{LC}(\theta,P)=\bigl(\frac{P}{30}\bigr)^{-1}\exp\bigl(-\frac{\theta^2}{2\theta_{0}^2}\bigr), \label{eq12}
\end{equation}
and $\theta$ is the pitch-angle measured from the sunward IMF direction, $P$ is the rigidity of GCRs in GV, $C_{LC}(t)$ ($\leq 0$) and $C_{EX}(t)$ ($\geq 0$) are amplitudes of the LC and EX anisotropies, respectively, and $\theta_{0}$ is the angle parameter denoting the width of the LC anisotropy. The $1/P$ dependence of the LC amplitude on $P$ is assumed to follow the average $P$-dependence of the size of the FD. It is seen in Figure 3 that the model anisotropy shown by a solid curve reproduces well the numerical pitch-angle distribution by paper 1.

By using $f$ defined in eq. (\ref{eq11}), we calculate the expected intensity $I^{cal}_{i(k,l)}(t)$ for the $(k,l)$ pixel in the $i$-th detector, as
\begin{eqnarray}
\hspace{1.0cm}
I^{cal}_{i(k,l)}(t)=I^{\prime 0}_{i(k,l)}(t) &+& \xi^{\prime GEO}_x(t)(c^{1}_{1 i(k,l)} \cos \omega t_i - s^{1}_{1 i(k,l)}\sin \omega t_i) \nonumber \\
 &+& \xi^{\prime GEO}_y(t)(s^{1}_{1 i(k,l)} \cos \omega t_i + c^{1}_{1 i(k,l)}\sin \omega t_i) + \xi^{\prime GEO}_z(t)c^{0}_{1 i(k,l)} \nonumber \\
   &+& C_{LC}(t)\frac{\int^{\infty}_{P^{cut}_{i(k,l)}}N_{i(k,l)}(P)f_{LC}(\theta_{i(k,l)}(P),P)dP}{\int^{\infty}_{P^{cut}_{i(k,l)}}N_{i(k,l)}(P)dP}, \label{eq13}
\end{eqnarray}
where $N_{i(k,l)}(P)$, representing the number of muons produced by primary particles with rigidity $P$ and recorded in the $(k,l)$ pixel in the $i$-th detector, is calculated by utilizing the response function of muons in the atmosphere to primary particles \citep{Murakami79} and $P^{cut}_{i(k,l)}$ represents the minimum (cut-off) rigidity of primary cosmic rays to produce muons recorded in the $(k,l)$ pixel. For detectors (Nagoya, Hobart and S\~{a}o Martinho) in which each $(k,l)$ pixel has a wide FOV resulting from the 2-fold coincidence between 1 m$^2$ unit detectors, we perform the integration in the last term for each combination of the virtual 0.1$\times$0.1 m$^2$ sub-detectors and then sum all values to get an integration. We perform only one integration for each $(k,l)$ pixel in Kuwait. The angular resolution of this integration is, therefore, 3.3$\degr$ for Nagoya, Hobart and S\~{a}o Martinho and 7.1$\degr$ for Kuwait. Note that the EX anisotropy, which is essentially the first order anisotropy along the IMF as described in eq. (\ref{eq11}), is included in $\xi^{\prime GEO}_x(t)$, $\xi^{\prime GEO}_y(t)$ and $\xi^{\prime GEO}_z(t)$, while the LC amplitude $C_{LC}(t)$ explicitly appears in the model intensity in eq. (\ref{eq13}). We will derive later $C_{EX}(t)$, therefore, by calculating the component of {\boldmath{$\xi$}}$^{\prime GEO}(t)$ parallel to the IMF. We repeat the calculation of $I^{cal}_{i(k,l)}(t)$ in eq. (\ref{eq13}) changing $\theta_{0}$ in every $1^{\circ}$ step and determine the best-fit parameters $C_{LC}(t)$, $I^{\prime }_{0}(t)$, {\boldmath{$\xi$}}$^{\prime GEO}(t)$ which minimize $S$ defined as
\begin{equation}
\hspace{1.0cm}
S=\sqrt{\frac{1}{M}\sum^{M}_{m=1} s^2(t_{m})}=\sqrt{\frac{1}{NM}\sum^{M}_{m=1}\sum^{N}_{i,k,l=1}\frac{(I^{obs}_{i(k,l)}(t_m)-I^{cal}_{i(k,l)}(t_m))^2}{\sigma^2_{i(k,l)}}},  \label{eq14}
\end{equation}
where $s(t_{m})$ is the hourly residual of the best-fitting at the time $t_{m}$ and $N$ and $M$ are the total number of directional channels and hours used for the best-fit calculations, respectively. Note that all the best-fit parameters $C_{LC}(t)$, $I^{\prime }_{0}(t)$, {\boldmath{$\xi$}}$^{\prime GEO}(t)$ at each time $t$ are uniquely determined by the linear least-square method for each value of $\theta_{0}$. We carry out this best-fit calculation for a total of 36 hours (DOYs 347.104$\sim$348.563) preceding the SSC onset and obtain the best-fit parameters with the minimum residual of $S=1.22$ and $\theta_{0}=35^{\circ}$. As we did for deriving $s I^{obs}_{i(k,l)}(t)$ in eq. (\ref{eq08}), we subtract $\bar{I}_{i(k,l)}(t)$ in eq. (\ref{eq02}) from $I^{cal}_{i(k,l)}(t)$ and get the reproduced intensity distribution $\Delta I^{cal}_{i(k,l)}(t)$, as
\begin{equation}
  \Delta I^{cal}_{i(k,l)}(t) = I^{cal}_{i(k,l)}(t)-\bar{I}_{i(k,l)}(t). \label{eq15}
\end{equation}
The 2D maps ($s I^{cal}_{i(k,l)}(t)$) reproduced from the best-fit parameters are shown in Figs.2b, 2d and 2f for S\~{a}o Martinho, Hobart and Kuwait, respectively. It is seen that both the LC and EX features are successfully reproduced with the best-fit parameters, although we don't see such a clear resemblance for Kuwait in Figure 2f due to the poor statistical significance of the data from this small muon hodoscope.  

Figure 4 displays the best-fit parameters, each as a function of time measured from the SSC. In Figure 4a, the hourly residual $s(t_{m})$ is shown by black circles, together with the residual in the conventional best-fit analyses in Figure 1 shown by gray circles. It is seen that the best-fitting is improved by the new model in eq. (\ref{eq13}) as indicated by the reduced residual, particularly when the best-fitting in Figure 1 fails with large residual due to the LC signatures observed by the GMDN. Black circles in Figure 4b display the LC amplitude ($C_{LC}(t)$) obtained for the LC periods when the zero pitch-angle is monitored by the GMDN, while gray circles display the amplitude obtained when the zero pitch-angle is out of the FOV of the GMDN. There is a large fluctuation seen in $C_{LC}(t)$ in this figure because we cannot make accurate best-fitting when the LC is out of the FOV. It is clear, however, that the LC amplitude ($C_{LC}(t)$) gradually decreases toward $\sim-6$ \% at the SSC onset. This LC amplitude is almost twice the FD size of $\sim$-3 \% (see Figure 1c), indicating that the maximum intensity depression behind the IP shock is much larger than the FD size recorded at the Earth. Also shown in the remaining panels are the best-fit parameters $I^{\prime }_{0}(t)$ and {\boldmath{$\xi$}}$^{\prime GEO}(t)$ transformed to the GSE coordinate system. In Figs.4d-4f, we plot parameters ($\Delta I_{0}(t)$, $\Delta \xi^{GSE}_x(t)$, $\Delta \xi^{GSE}_y(t)$, $\Delta \xi^{GSE}_z(t)$) respectively calculated, as
\setlength{\arraycolsep}{0.0em}
\begin{eqnarray}
  \Delta I_{0}(t) = I^{\prime }_{0}(t)-\bar{I}_{0}(t), \label{eq16} \\
  \Delta \xi^{GSE}_x(t) = \xi^{\prime GSE}_x(t)-\bar{\xi}^{GSE}_x(t), \label{eq17} \\
  \Delta \xi^{GSE}_y(t) = \xi^{\prime GSE}_y(t)-\bar{\xi}^{GSE}_y(t), \label{eq18} \\
  \Delta \xi^{GSE}_z(t) = \xi^{\prime GSE}_z(t)-\bar{\xi}^{GSE}_z(t), \label{eq19}
\end{eqnarray}
after subtracting the 24 hour TMAs of the conventional best-fit parameters representing the contribution from the diurnal anisotropy. We note $\Delta \xi^{GSE}_x(t)$ and $\Delta \xi^{GSE}_y(t)$ in Figs.4d and 4e showing positive and negative deviations from zero, respectively, during $\sim$10 hours preceding the SSC. This is consistent with the enhancement of the excess intensity from the sunward IMF direction, as expected from the measurement of the shock reflected particles. This is confirmed in Figure 4g showing the anisotropy from the sunward IMF direction. This parallel anisotropy shows the maximum deviation of $\sim$1 \% at the SSC onset which is consistent with the expectation from the measurement of the shock reflected particles as discussed in eq. (\ref{eq10}). We finally note the density ($\Delta I_{0}(t)$) also showing a positive deviation similar to the parallel anisotropy. The maximum deviation is $\sim$1 \% at 0 hour from the SSC and is almost the same as the maximum deviation of the parallel anisotropy, giving a support to our model function of the EX anisotropy in eq. (\ref{eq11}), wherein the second term of the right hand side contains two terms independent and dependent on the pitch-angle ($\theta$) with the same amplitude at $\theta=0^{\circ}$. In the next section, we will compare these best-fit parameters with the theoretical predictions by paper 1 and discuss their physical implications. 

\section{Summary and discussions}  

We have analyzed cosmic ray precursors of a Forbush Decrease (FD) observed on December 2006 with the GMDN monitoring the directional intensity of $\sim$50 GeV galactic cosmic rays. This Forbush Decrease was caused by an IP shock associated with a CME which erupted from the Sun shortly after an X3.4 flare on December 13 and arrived at the Earth in December 14. An intense geomagnetic storm was also recorded during this FD with the peak Kp index of 8+. There was no other SSC and/or interplanetary disturbances recorded at the Earth during a period between the onsets of the flare and the SSC, which is analyzed in this paper. By using the 2D maps of the cosmic ray intensity produced by removing the contribution from the diurnal anisotropy, we found a clear signature of the LC anisotropy which was observed as a deficit intensity from the sunward IMF direction. The significant LC signature was first recorded by Hobart at $\sim20$ hours before the SSC and then by S\~{a}o Martinho with a larger amplitude at $\sim-6$ hours. A weak LC signature was also seen in Kuwait viewing the sky in between FOVs of Hobart and S\~{a}o Martinho.

The ``half-width'' opening angle $\theta_{HW}$ of the LC is defined in paper 1 as the pitch-angle at which the intensity decrease (relative to the omnidirectional intensity) has reached half its maximum value. According to this definition, $\theta_{HW}$ is calculated to be $36.4^\circ$ using $\theta_{0}=35^{\circ}$ obtained in \S 3.2. By linearly interpolating the numerical relationship between the loss-cone width ($\theta_{HW}$) and the angle between the shock normal and the upstream magnetic field ($\theta_{Bn}$) given by paper 1 (see Table 2 in paper 1), we find $\theta_{HW}=36.4^\circ$ corresponding to $\theta_{Bn}= 23^\circ$ assuming the ``local'' slope of the power spectrum of the IMF turbulence ($q=0.5$) appropriate for the muon detector. This $\theta_{Bn}$ is less than half of $\theta_{Bn}=56^{\circ}$ obtained by \citet{Liu08}(see \S 2). One possible source of this discrepancy is that a large error might associate with the non-linear best-fit parameter $\theta_{0}$ obtained in \S 3.2. If we evaluate the error by calculating $\theta_{0}$ with which $S^{2}$ reaches at $S^{2}=2.49=1.22^{2}+1$, that is, +1.0 more than $S^{2}$ with minimum $S$, $\theta_{0}$ with the error is estimated to be $\theta_{0}=35^{+15^{\circ}}_{-12^{\circ}}$ which corresponds to $\theta_{Bn}=23^{+30^{\circ}}_{-13^{\circ}}$. This $\theta_{Bn}$ with a large error is not inconsistent with $\theta_{Bn}=56^{\circ}$ by \citet{Liu08}. We are not sure whether this error estimation is appropriate or not, but we also carried out the best-fit calculation in \S 3.2 with $\theta_{0}$ fixed at 23$\degr$ corresponding to $\theta_{Bn}=56^{\circ}$ and confirmed that the minimum residual $S$ and the best-fit parameters except $\theta_{0}$ are almost unchanged from those in Figure 4. The reproduced 2D maps were also very similar to those in Figs.2b, 2d and 2f. This implies that $\theta_{Bn}$ cannot be determined accurately by the present analysis method for the GMDN data. One reason for such a large error in $\theta_{Bn}$ (or $\theta_{0}$) is the poor angular resolution of the observed incident direction in detectors (Hobart and S\~{a}o Martinho) with which the LC anisotropy was recorded. This is different from another LC event on October 2003 which was observed with a large muon hodoscope with the better angular resolution \citep{kmuna05}. It is also noted that the dependence of $\theta_{HW}$ (i.e. $\theta_{0}$) on $\theta_{Bn}$ becomes very small for $\theta_{Bn}$ above $\sim30^\circ$ according to paper 1. This means that one has to measure $\theta_{0}$ very accurately for precisely determining $\theta_{Bn}$ particularly when $\theta_{Bn}$ is large in case of the quasi-perpendicular shock. 

A striking feature of this event is that a weak LC signature was also recorded by S\~{a}o Martinho more than a day earlier on December 13 at $\sim-28$ hours. To examine the lead time of this LC precursor, we fit to $C_{LC}(t)$ in Figure 4b an exponential function of time $t$ (measured from the SSC) defined as,
\begin{equation}
C_{LC}(t)=C_{LC}(0) \exp \bigl(\frac{t}{T_{0}}\bigr), \label{eq20}
\end{equation}
and obtain $C_{LC}(0)=-6.45$ \% and $T_{0}=15.6$ hours. For this fitting, we used only $C_{LC}(t)$ obtained when the zero pitch-angle is monitored by the GMDN (as displayed by solid circles in Figure 4b). This lead time $T_{0}$ for 30 GeV particles corresponds to the ``decay length'' $l$ of the LC, as
\begin{equation}
l=T_{0}V_{S}/\cos \theta_{Bn} \sim 0.42 \rm{AU}, \label{eq21}
\end{equation}
with $\theta_{Bn}=23^{\circ}$ and $V_{S}=1030$ km/s \citep{Liu08}. By using a ratio $l/\lambda=0.16$ derived for $q=0.5$ and $\theta_{HW}=36.4^{\circ}$ from interpolating Table 2 in paper 1, we estimate the parallel mean free path for the pitch-angle scattering ($\lambda$) to be $\sim$2.6 AU, which is almost two times longer than 1.5 AU estimated by paper 1 for the muon detector. This indicates that the IMF was more quiet in this event than a typical power spectrum assumed for the IMF turbulence by paper 1. These conclusions with $q=0.5$ remain unchanged even if we choose $q=1.0$, with which we get $\theta_{Bn} \sim 25^{\circ}$, $l \sim 0.43$ AU and $\lambda \sim 3.2$ AU.

We have also found the excess intensity from the sunward IMF direction clearly observed during $\sim$10 hours preceding the SSC. We have shown that this excess intensity is consistent with the measurement of the particles accelerated by the head-on collisions with the approaching shock. This is the first detailed observation of the precursor due to the shock reflected particles with muon detectors. It is interesting to note that there is also an enhancement seen in the anisotropy perpendicular to the IMF in Figure 4h. We think that this perpendicular anisotropy is probably arising from the drift anisotropy expressed by the vector product between the IMF and the spatial gradient of the cosmic ray density, which directs toward the shock due to the higher population of the reflected particles at the shock. By using the observed perpendicular anisotropy together with the IMF data, therefore, we can deduce the temporal variation of the cosmic-ray density gradient and infer the geometry of the IP shock. In our separate papers, we actually succeeded in deriving the geometry of the magnetic flux rope in ICMEs by this method \citep{Kuwa04, Kuwa09}.

The precursor observed with the long lead time, like the event analyzed in this paper, is of particular importance for the possible space weather forecast using cosmic ray measurements. For the accurate observation of such event, however, we need further improvement of the GMDN. First, the incomplete sky-coverage of the GMDN allowed us to analyze the best-fit parameters during only half a period in Figure 4, when the sunward IMF direction was in the FOV of the GMDN. Second, the insufficient detection areas of Hobart and Kuwait increased the statistical error and introduced a non-uniformity into the response of the GMDN to the LC precursor. We believe that such a non-uniform response also contributes to the large fluctuations in the best-fit parameters in Figure 4. We are now planning to overcome these technical problems by expanding the detection areas of Hobart and Kuwait and also by installing new detector(s) to expand the FOV of the GMDN preparing for the next solar maximum expected at around the year 2013.

\acknowledgments

This work is supported in part by NASA grant NNX 08AQ01G, and in part by Grants-in-Aid for Scientific Research from the Ministry of Education, Culture, Sports, Science and Technology in Japan and by the joint research programs of the Solar-Terrestrial Environment Laboratory, Nagoya University. The observations with the Kuwait Muon Telescope are supported by the Kuwait University grant SP03/03. We thank N. F. Ness for providing {\it ACE} magnetic field data via the {\it ACE} Science Center.

\clearpage



\figcaption[fig1]{Hourly mean solar wind and cosmic ray data observed during three days between 13 and 15 December, 2006. Each panel from the top shows (a) the {\it ACE}-Level2 solar wind velocity, (b) the {\it ACE}-Level2 IMF magnitude, (c) the best-fit GCR density ($I_{0}(t)$), (d-f) three components ($\xi^{GSE}_{x}(t)$, $\xi^{GSE}_{y}(t)$, $\xi^{GSE}_{z}(t)$) of the GCR anisotropy observed by the GMDN and (g) the pressure-corrected count rate ($I^{obs}_{i(k,l)}(t)$) recorded in the vertical channel of the S\~{a}o Martinho detector in Brazil, as a function of time in the day of year (DOY) on the horizontal axis. The anisotropy components in this figure are transformed to the GSE coordinate system and corrected for both the solar wind convection anisotropy and the Compton-Getting anisotropy arising from the Earth's orbital motion around the Sun. The onset times of the X3.4 flare at 02:14 UT on December 13 and the SSC at 14:14 UT on December 14 are indicated by vertical solid lines. The gray curves in c-f display the 24 hour TMAs of data shown by solid circles ($\bar{I}_{0}(t)$, $\bar{\xi}^{GSE}_x(t)$, $\bar{\xi}^{GSE}_y(t)$, $\bar{\xi}^{GSE}_z(t)$, in eqs.(3)-(6) ). The gray curve in g shows $\bar{I}_{i(k,l)}(t)$ for the vertical channel of S\~{a}o Martinho expected from the 24-hour TMAs of the best-fit parameter (see \S 3.1 in text), while open circles in g display $\Delta I^{obs}_{i(k,l)}(t)$ (see text). The gap in the solar wind velocity data between 05:00-18:00 UT on December 13 (DOYs 347.229-347.354) is filled by using the {\it ACE}-Level2 alpha-particle velocity.  \label{fig01}} 
\clearpage

\figcaption[fig2]{2D significance maps observed by the GMDN on December 13 and 14 preceding the SSC onset. Panels a, c and e display the 2D maps ($s I^{obs}_{i(k,l)}(t)$) observed by S\~{a}o Martinho, Hobart and Kuwait, respectively, while b, d and f show the maps ($s I^{cal}_{i(k,l)}(t)$) reproduced from the best-fit parameters obtained in \S 3.2. Each small square panel in this figure displays the ``significance'' in one hour in a color-coded format as a function of $k$ and $l$ on the horizontal and vertical axes as indicated at the left and bottom edges of each figure. Red (blue) color denotes the excess (deficit) intensity in ($k,l$) pixel relative to the ominidirectional intensity in the FOV. The color scale is set $\pm$ 5 for S\~{a}o Martinho and $\pm$ 3 for Hobart and Kuwait as indicated by color bars at the right bottom corners of panels b and d. Also shown by the white curve in each square panel is the contour line of the pitch-angle measured from the sunward IMF direction and calculated for cosmic rays incident to each ($k,l$) pixel with the median primary energy appropriate to that pixel. All maps during 36 hours are shown in a-d for S\~{a}o Martinho and Hobart, while maps by Kuwait are shown only for 6 hours when the weak LC signature is visible with this small detector.  \label{fig02}}
\clearpage

\figcaption[fig3]{Model pitch-angle distribution for the best-fit analyses. Open circles display the numerical GCR intensity read from FIG.5 of \citet{Lee03} as a function of the pitch-angle cosine on the horizontal axis, while the solid line shows our model distribution which is the sum of the Gaussian (plus signs) and pitch-angle cosine (crosses) functions denoting the LC and the EX anisotropies, respectively. In this figure, the model distribution is normalized to 1.239 at $\mu=-1$.  \label{fig03}}
\clearpage

\figcaption[fig4]{Best-fit parameters in the model pitch-angle distribution. These parameters are obtained from the best-fitting to the observed 2D maps in Figure 2. Each panel from the top displays by solid circles (a) the hourly residual ($s(t_{m})$), (b) the amplitude of the LC anisotropy ($C_{LC}(t)$), (c) the density ($\Delta I_{0}(t)$), (d-f) three components of the first order anisotropy in the GSE coordinate system ($\Delta \xi^{GSE}_x(t)$, $\Delta \xi^{GSE}_y(t)$, $\Delta \xi^{GSE}_z(t)$) after subtracting 24-hour TMA, (g) and (h) the anisotropy component parallel and perpendicular to the sunward IMF direction, respectively, as a function of time measured from the SSC. For comparison, the residual in the conventional best-fit analysis is also plotted by gray circles in a. Solid circles in b-h display the parameters obtained when the sunward IMF direction is monitored by the GMDN, while gray circles display parameters obtained when the sunward IMF direction is out of the FOV of the GMDN. Each gray curve in b, c and g is the best-fit to the solid circles with the exponential function of the time (see text).  \label{fig04}}
\clearpage

\begin{figure}
  \figurenum{1}
  \caption{}
  \centering
  \includegraphics[width=5.0in]{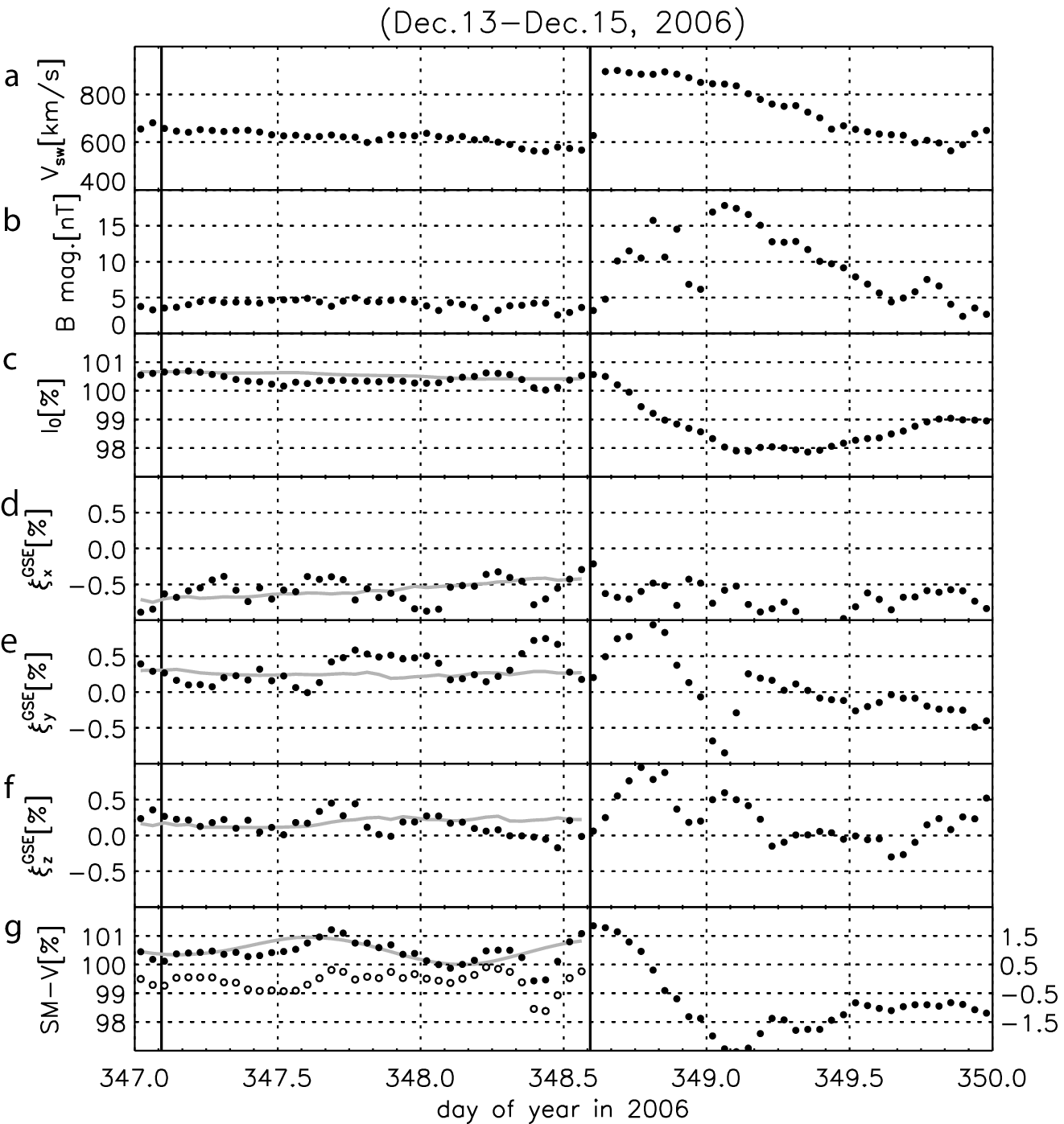}
\end{figure}
\clearpage

\begin{figure}
  \figurenum{2}
  \caption{}
  \centering
  \includegraphics[width=6.5in]{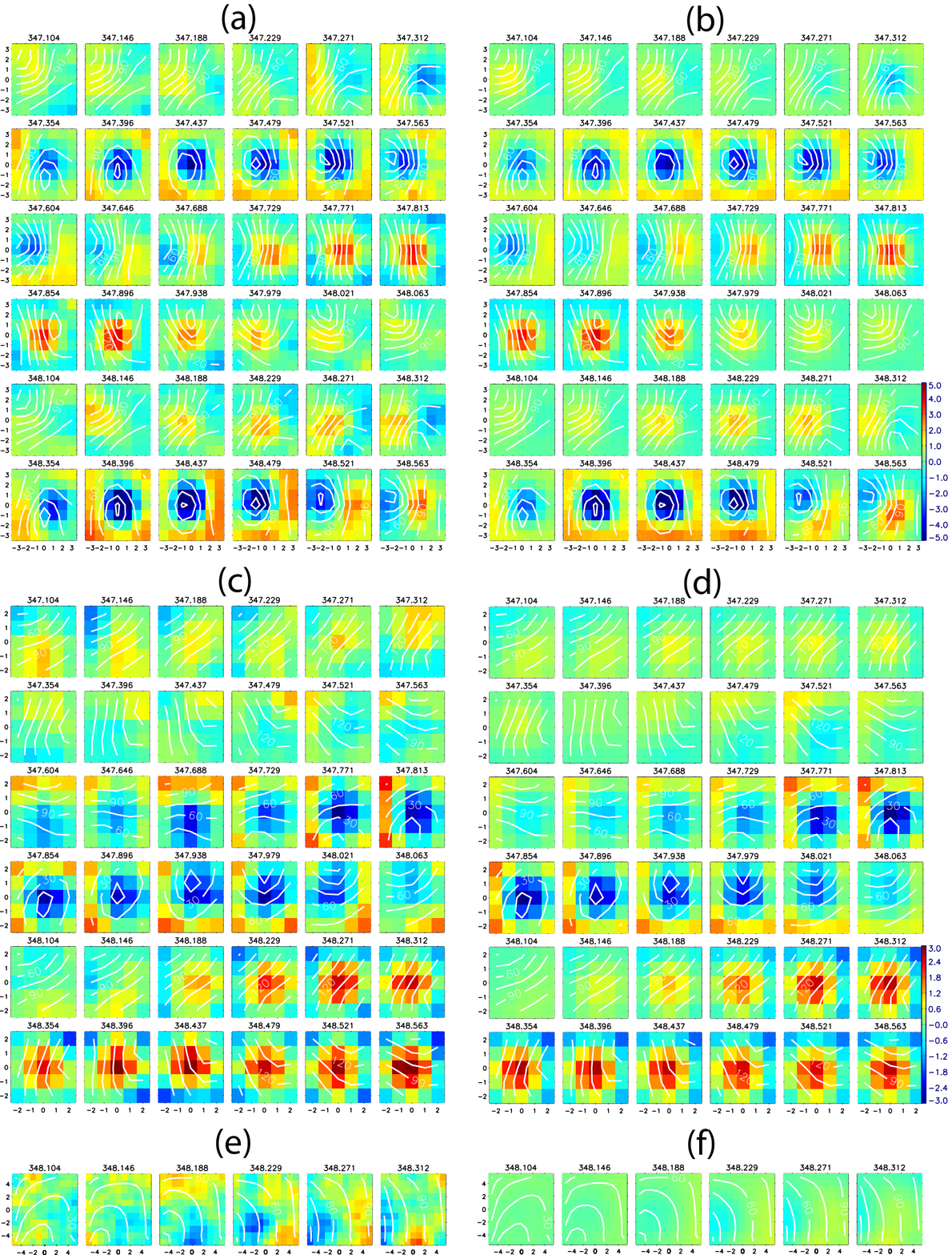}
\end{figure}
\clearpage

\begin{figure}
  \figurenum{3}
  \caption{}
  \centering
  \includegraphics[width=4.0in]{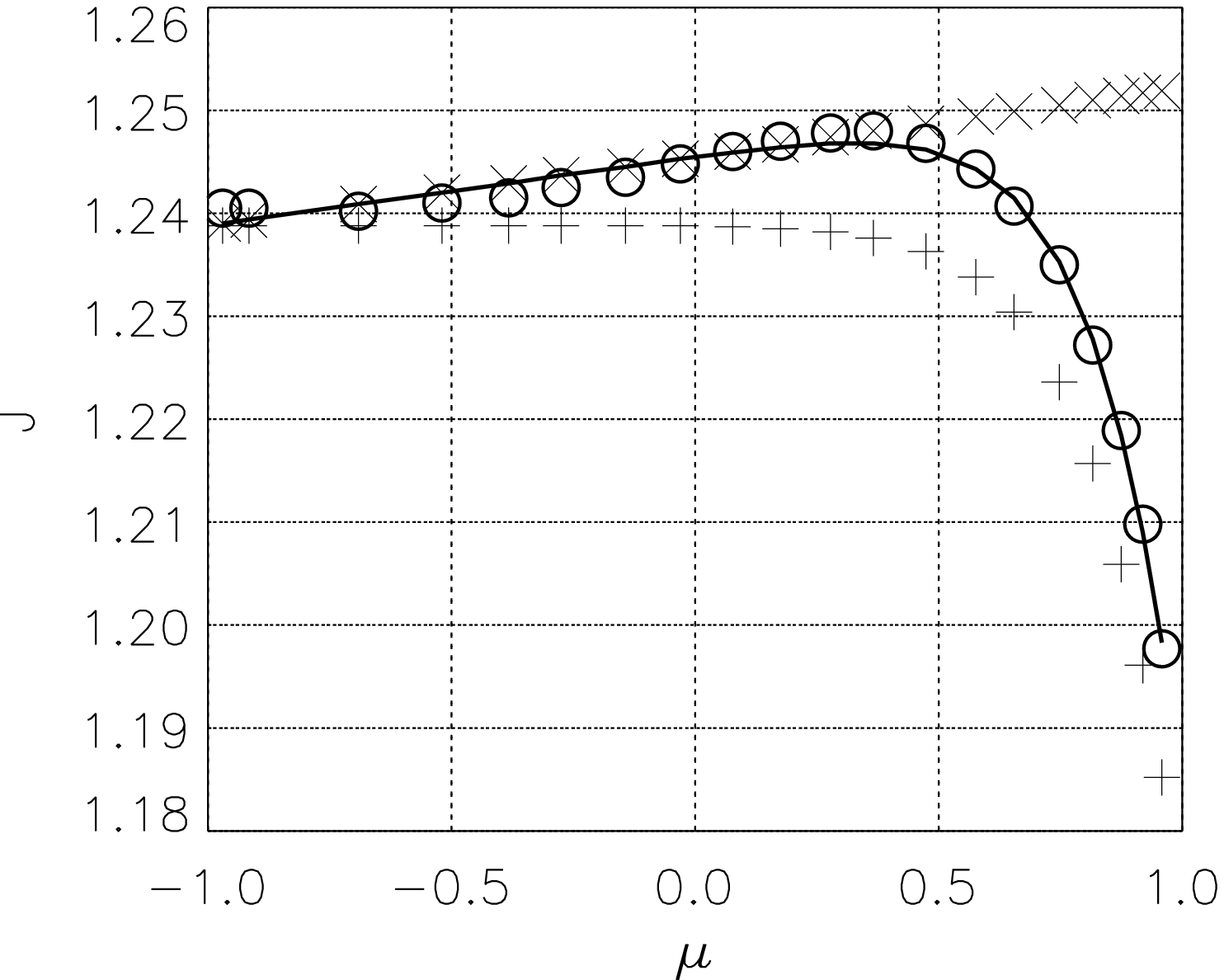}
 \end{figure}
\clearpage

\begin{figure}
  \figurenum{4}
  \caption{}
  \centering
  \includegraphics[width=5.0in]{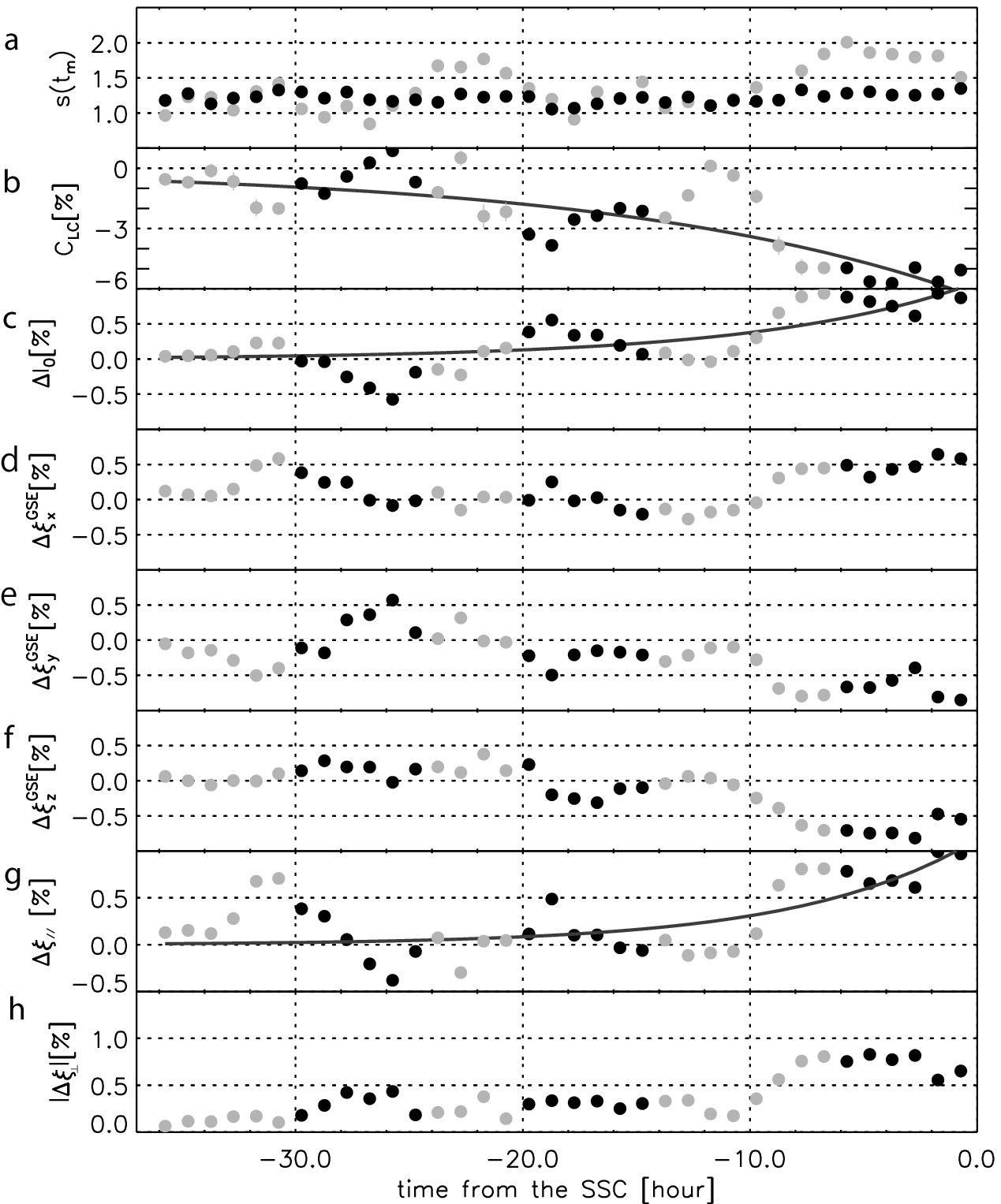}
\end{figure}

\end{document}